\documentstyle[epsfig,latexsym,amssymb,aps,floats,eqsecnum,preprint,
amsfonts]{revtex}
\def\laq{\raise 0.4ex\hbox{$<$}\kern -0.8em\lower 0.62 ex\hbox{$\sim$}}
\def\gaq{\raise 0.4ex\hbox{$>$}\kern -0.7em\lower 0.62 ex\hbox{$\sim$}}

\begin{document}
\begin{titlepage}
\begin{flushright}
CERN-TH/2003-017
\end{flushright}
\vspace*{1cm}

\begin{center}
{\large{\bf Low-scale Quintessential Inflation}}
\vskip 1 cm 
{\sl Massimo Giovannini\footnote{Electronic Address: massimo.giovannini@cern.ch}}
\vskip 0.5 cm 
{\sl  Theoretical Physics Division, CERN,CH-1211, Switzerland}
\vspace{1cm}
\noindent
\begin{abstract}
In quintessential inflationary model, 
the same master field that drives inflation becomes, 
later on, the dynamical source of the (present) accelerated expansion. 
Quintessential inflationary models require a curvature scale at the end of inflation 
around $10^{-6}M_{\rm P}$ in order to explain the 
large scale fluctuations observed in the microwave sky.
If the curvature scale at the end of inflation
is much smaller than $10^{-6}M_{\rm P}$,
the large scale adiabatic mode may be produced thanks to the relaxation 
of a scalar degree of freedom, 
which will be generically denoted, according to the recent terminology, 
as the curvaton field.  
The production of the adiabatic 
mode is analysed in detail in the case of the minimal quintessential
inflationary model originally proposed by Peebles and Vilenkin. 
\end{abstract} 
\vskip0.5pc
\end{center}
\end{titlepage}
\newpage
\noindent

\renewcommand{\theequation}{1.\arabic{equation}}
\setcounter{equation}{0}
\section{Introduction} 
Following the discovery \cite{rie,per} that distant supernovae are fainter 
than inferred from local samples, models of scalar fields that are able to 
develop a negative pressure at the present time have been 
proposed. These scalar fields only interact gravitationally and 
they have been generically named quintessence \cite{rat,stei0,stei1,stei2,car}. 

In recent years, an interesting class of quintessence 
models have been proposed by Peebles and Vilenkin \cite{pv1} 
(see also \cite{pv2}). The peculiar feature of quintessential inflation is 
that the  master field that drove inflation becomes again dominant, 
at the present time. Hence, in the context of quintessential inflation, 
the inflaton and the quintessence field are identified in a single 
scalar degree of freedom, driving inflation in the past and 
acting as quintessence today. A simple example of this dynamics 
is represented by ``dual'' potentials \cite{pv1} going as a power {\em during} 
inflation and as an inverse power {\em after} \cite{rat} inflation. 
Owing  to this dual form of the potential,the dynamics of the background
 will experience, right after inflation, 
a pretty long phase in which the potential term is subleading  with
respect to
 the kinetic term. In quintessential inflation the reheating can 
be entirely gravitational, since the energy density in 
 the light quanta produced at the end of inflation decreases slower than the
(kinetic-energy-dominated) background geometry \cite{pv1,spok}. 

Quintessential inflationary models are consistent with observations provided 
the curvature scale at the end of inflation is not too low and  
${\cal O} \sim (10^{-6}M_{\rm P})$. In the present paper a complementary 
possibility will be analysed. Consider the situation where 
$H_{\rm e}$, the curvature scale at the end of inflation, is indeed much 
smaller than $10^{-6}M_{\rm P}$. Along this line we suppose, for simplicity,
 that during inflation there is degree 
of freedom, $\psi$, which is not coupled with the inflaton field and 
remains constant, thanks to its potential, during the  later 
stages of inflation. The (potential) energy density 
of $\psi$ is subleading  with resepect to the energy density of $\varphi$. 
At the end of inflation, during  the kinetic phase, the large scale 
fluctuations of $\psi$ will be converted into adiabatic fluctuations. 
This  possibility has  recently  been studied by many authors 
\cite{es,lw,us,sl,mor,bas,gor}
in different contexts, and it was originally invoked in \cite{mol}.
As discussed in  \cite{bart}, even the simplest chaotic 
inflationary models develop new constraints when combined with the curvaton 
idea. The purpose of the present paper is to analyse 
low-scale quintessential inflation in a specific set-up, which is the 
one originally suggested by Peebles and Vilenkin in \cite{pv1}.
In this model the late-time behaviour of the quintessential evolution does not show the 
tracking behaviour of the inflaton and matter mass densities, as argued 
in \cite{stei2}. 

Particular attention will be given to the evolution of the 
fluctuations. This is quite essential since, by lowering 
the curvature scale at which inflation ends, we have to make sure 
that the isocurvature mode of $\psi$ is efficiently turned into an 
adiabatic one. While in ordinary inflationary models (such as those analysed, 
for instance, in \cite{bart}) the radiation-dominated 
phase starts at the end of inflation, in quintessential inflation
the evolution may be very different and the onset of the 
radiation-dominated phase is delayed.  One of the purposes of the present paper is 
indeed to generalize the analysis of the curvaton evolution 
to backgrounds where inflation is not immediately followed 
by a radiation-dominated phase. The curvature perturbation will be followed through 
all the stages of the model and its final value computed. This analysis 
will be performed both analytically and numerically. 

The plan of the present  paper is the following. In Section II the 
constraints on the post-inflationary evolution will be derived in the 
specific case where the inflaton and quintessence field are identified.
In Section III the constraints pertaining to the quintessential 
evolution will be scrutinized. Section IV contains the basic 
ingredients for the evolution of the fluctuations.
Sections V and VI deal with the conversion of the initial
isocurvature mode into an adiabatic one. In Section V the initial and the kinetic stages 
will be scrutinized, while Section VI is more oriented towards the phase where $\psi$ dominates and 
eventually decays. In Section VII some concluding remarks will be presented.

\renewcommand{\theequation}{2.\arabic{equation}}
\setcounter{equation}{0}
\section{From inflation to quintessence} 
Consider the  minimal  realization of a low-scale quintessential 
inflationary model extending the original proposal of \cite{pv1}:
\begin{eqnarray}
&& M_{\rm P}^2 H^2 = \biggl[ \frac{\dot{\varphi}^2}{2} + \frac{\dot{\psi}^2}{2}
+ V(\varphi) + W(\psi)\biggr],
\label{b1}\\
&& M_{\rm P}^2 ( H^2 + \dot{H}) = \biggl[ - \dot{\varphi}^2 - \dot{\psi}^2
+ V(\varphi) + W(\psi) \biggr],
\label{b2}\\
&& \ddot{\varphi} + 3 H \dot{\varphi} + \frac{\partial V}{\partial \varphi} =0,
\label{phb}\\
&& \ddot{\psi} + 3 H \dot{\psi} + \frac{\partial W}{\partial\psi} =0.
\label{psb}
\end{eqnarray}
The potential of $\varphi$ can be chosen to be 
a typical power law during inflation and an {\em inverse} power  during 
the quintessential regime:
\begin{eqnarray}
&&V(\varphi) = \lambda ( \varphi^4 + M^4) ,~~~~ \varphi < 0,
\nonumber\\
&& V(\varphi) = \frac{\lambda M^8}{\varphi^4 + M^4}, ~~~~ \varphi \geq 0,
\label{potph}
\end{eqnarray}
where $\lambda$ is the inflaton self-coupling and  $M$ is the typical 
scale of quintessential evolution.

The field $\psi$ is subleading during inflation and it is characterized by 
a potential, which we will take, for simplicity, to be  quadratic, i.e. 
\begin{equation}
W(\psi) = \frac{m^2}{2} \psi^2.
\label{potps}
\end{equation}
This set-up can be generalized to the case where 
the field $\psi$ is replaced by an arbitrary number of scalar 
degrees of freedom $\psi_{i}$.

Inflation ends when $V(\varphi) \sim \lambda M_{\rm P}^4$ at a 
curvature scale $H_{\rm e}$ 
\begin{equation}
\sqrt{\lambda} \simeq \frac{H_{\rm e}}{M_{\rm P}}.
\end{equation}  
In ordinary inflationary models, the large scale 
inhomogeneities determining the CMB anisotropies come from 
the Gaussian fluctuations of the inflaton $\varphi$, and, consequently, 
$\lambda\sim 10^{-13}$ \cite{pv1}. In the present investigation a complementary 
possibility will be discussed, namely the case when 
\begin{equation}
 H_{\rm e} \ll 10^{-6} M_{\rm P}.
\label{lows}
\end{equation}
In this case the fluctuations of $\varphi$ are too small to be 
interesting for CMB physics.
However, this conclusion can be evaded by taking into account 
the fluctuations of the field $\psi$. Qualitatively the 
picture is the following. Right after inflation, the fluctuations 
of the geometry will be determined 
by the fluctuations both of $\varphi$ and 
$\psi$. Because of Eq. (\ref{lows}), the metric fluctuations 
generated by $\varphi$ will vanish at $t_{\rm e}$. In other
words, the initial conditions for the system at $t_{\rm e}$ will be of 
isocurvature type. Later on, the isocurvature mode can be converted into an adiabatic one.
This is the idea explored, for instance, in \cite{es,lw}. In the usual 
picture of curvaton evolution, right after inflation, radiation takes place immediately. 
In the case of quintessential inflation, on the contrary, the onset of the radiation-dominated 
epoch may be delayed. Hence, the analysis of the evolution of the fluctuations must be 
repeated, with particular attention to the different features of the model. 

During inflation, the field $\psi$ is  subdominant with respect  to the inflaton 
energy density,
\begin{equation}
W(\psi) \ll V(\varphi).
\label{sub}
\end{equation}
The field $\psi$ remains nearly constant during the later stages 
of inflation, 
i.e. $\psi \sim \psi_{\rm e}$.
At the end of inflation, Eq. (\ref{sub}) implies that 
\begin{equation}
\biggl( \frac{m}{M_{\rm P}} 
\biggr) \ll \sqrt{2} \biggl( \frac{H_{\rm e}}{\psi_{\rm e}}\biggr).
\label{con1}
\end{equation}

After the end of inflation, because of the inverse power-law form of the potential,
the field $\varphi$ is mainly driven by its kinetic energy and the 
approximate solution 
of the background geometry will be of the type 
\begin{equation}
\varphi = \sqrt{2} M_{\rm P} \ln{\biggl( \frac{a}{a_{\rm e}}\biggr)},
\label{vfkin}
\end{equation}
with $a(t) \sim t^{1/3}$. In this 
phase the potential,
$V(\varphi) \sim \lambda (M/M_{\rm P})^4 \ln^{-4}{(a/a_{\rm e})}$, 
is subleading since $\lambda \ll 10^{-13}$ and $M \ll M_{\rm P}$.
More specifically, phenomenological considerations related to the present dominance 
of $\varphi$ (see Section III) suggest that $M \ll 10^{-9} M_{\rm P}$.

While the background is dominated by $\dot{\varphi}^2$, the field $\psi$ 
slightly decreases with a rate given by $W_{,\psi}$. From Eq. (\ref{psb}), the field $\psi$ 
slowly rolls toward the minimum of its potential in a kinetic-energy-dominated environment
and its approximate equation obeys
\begin{equation}
\dot{\psi} \simeq - \frac{1}{6 H} \frac{\partial W}{\partial \psi},
\label{psap}
\end{equation}
where the factor $1/6$ comes from dropping consistently the terms of Eq. (\ref{psb}) 
containing more than one derivative of the potential. This type of approximation 
has been exploited in \cite{us}, but in the case of a slow-roll occurring during the 
radiation epoch.

Using Eq. (\ref{psap}), and recalling that  $a(t)\sim t^{1/3}$,
it can be checked that the  evolution of $\psi$ slightly 
deviates from a constant value. For instance, in the case of a quadratic potential the 
solution of Eq. (\ref{psap}) can be written as 
\begin{equation}
\psi(t) \simeq \psi_{\rm e} \biggl[ 1 - \frac{m^2}{4} (t^2 - t_{\rm e}^2)\biggr].
\end{equation}
Equation (\ref{psap}) is useful in the case when the potential 
is more complicated than the quadratic ansatz of Eq. (\ref{potps}) where 
Eq. (\ref{psb}) can be solved exactly.
\begin{figure}
\begin{center}
\begin{tabular}{|c|c|}
      \hline
      \hbox{\epsfxsize = 7.5 cm  \epsffile{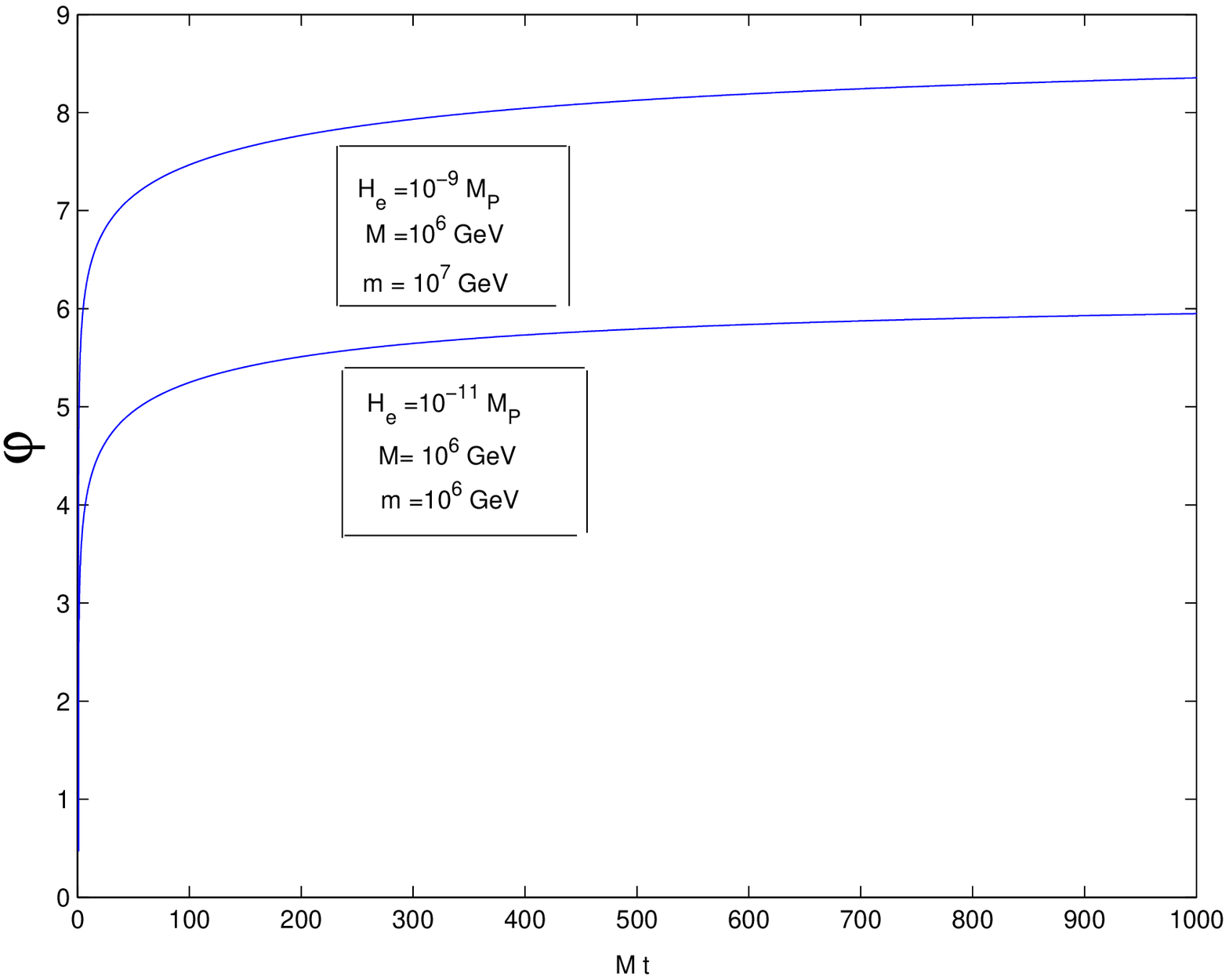}} &
      \hbox{\epsfxsize = 7.5 cm  \epsffile{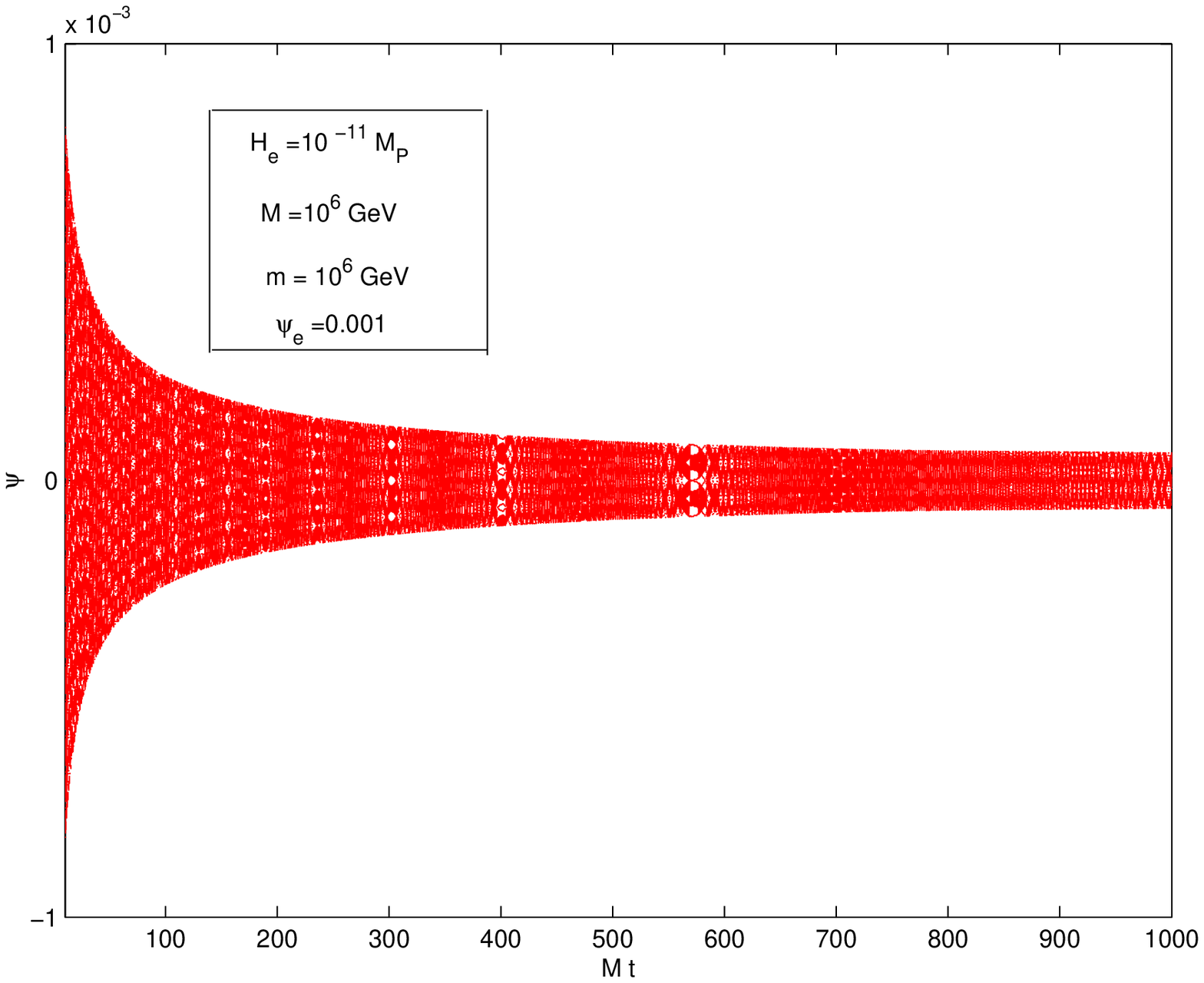}}\\
      \hline
\end{tabular}
\end{center}
\caption[a]{The result of the numerical integration of the background are illustrated for 
different values of the parameters (indicated above each curve).}
\label{varphevol}
\end{figure}
In fact, during the kinetic regime the curvature scale decreases, and when 
$H\sim H_{\rm osc} \sim m$ the field $\psi$, still subdominant, 
will start oscillating. The case of massive potential  
allows analytical solutions of Eq. (\ref{psb}) 
 during the kinetic phase for $\psi$:
\begin{equation}
\psi = \psi_{\rm e} \frac{\sqrt{t}}{a^{3/2} J_{0}(m t_{\rm e})} J_{0}(m t), 
\label{ansolpsi}
\end{equation}
which leads to Eq. (\ref{psap}) in the limit $ m t \ll 1$ and $m t_{\rm e} \ll 1$.
Since $\psi$ decays as $a^{-3/2}$ during the oscillating regime, 
it will become dominant with respect to the kinetic energy of $\varphi$ 
at a scale
\begin{equation}
H_{\rm d} \sim 5~ m \biggl(\frac{\psi_{\rm e}}{M_{\rm P}}\biggr)^2. 
\label{con2a}
\end{equation}
The field $\psi$ eventually decays at a curvature scale 
\begin{equation}
H_{\rm r} \sim \frac{m^3}{M_{\rm P}^2}. 
\label{con2b}
\end{equation}
At $H_{\rm m}$ the potential of $\varphi$ is still subdominant with respect 
to the kinetic energy. In fact 
\begin{equation}
V(\varphi_{\rm m}) \sim \lambda M^4 \biggl( \frac{M}{M_{\rm P}}\biggr)^4 
\ln^{-4}{\biggl(\frac{H_{\rm e}}{m}\biggr)}\ll \frac{ \dot{\varphi}^2_{\rm m}}{2} 
\simeq \frac{1}{9} \biggl(\frac{M_{\rm P}^2}{m^2} \biggr),
\end{equation}
where the last equality comes from  the kinetic term  of $\varphi$
at $t_{\rm m}$, evaluated on the basis of Eq. (\ref{vfkin}).
An example of the numerical integration is reported in Fig. \ref{varphevol}.
Furthermore, the analytical evolution of $\psi$, as reported in Eq. (\ref{ansolpsi}), 
is in excellent agreement with the numerical results.

Between $H_{\rm d}$ and $H_{\rm r}$,
\begin{equation}
\biggl(\frac{a_{\rm d}}{a_{\rm r}}\biggr) \simeq 
\biggl(\frac{m}{\psi_{\rm e}}\biggr)^2,
\end{equation}
 the background geometry is 
effectively dominated by the oscillations 
of $\psi$, i.e. $a(t) \sim (m t)^{2/3}$.

\begin{figure}
\centerline{\epsfxsize = 11cm  \epsffile{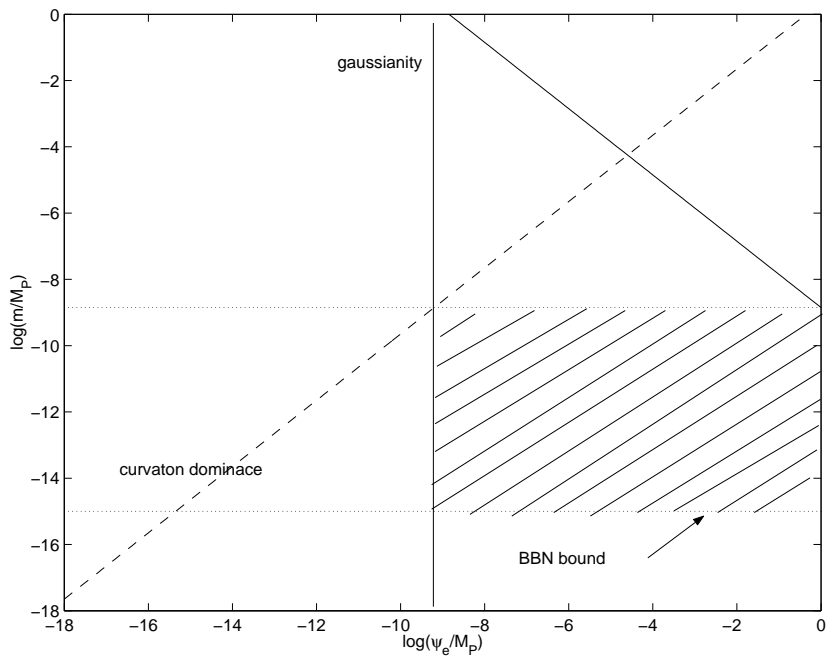}}
\vskip 3mm
\caption[a]{The quantitative constraints pertaining to the combined 
analysis of the evolution of $\psi$ and $\varphi$ during the kinetic phase 
are illustrated. With the dashed line the constraint coming from the 
dominance of $\psi$ is reported. The dot-dashed lined (from top to bottom) 
represent, respectively, the bound (\ref{con3}) and the BBN bound. The full
(vertical) line refers to Eq. (\ref{con4}). The shaded region defines the allowed
portion of the parameter space of the model.}
\label{constraints} 
\end{figure}

It is interesting to combine in the physical 
picture the constraints and the requirements 
introduced so far.
According to Eqs. (\ref{con2a}) and (\ref{con2b}) the quintessence 
field may become subdominant either prior to or after the decay of $\psi$.
In Fig. \ref{constraints}, with the (diagonal) dashed line, the 
condition coming from the interplay between the decay of $\psi$ and 
the dominance of $\varphi$ is illustrated for the specific case $ 
H_{\rm e} \sim 10^{-9} M_{\rm P}$. Above the dashed 
line, $\psi$ decays during the kinetic phase. Below, the dashed line 
the decay occurs when $\psi$ already dominates. The requirement 
of Eq. (\ref{con1}) also imposes  a constraint on Fig. \ref{constraints}, 
implying that $\psi_{\rm e}$ and the mass should lie below the full 
fine in the right-hand corner.
Consider now the situation where the fluctuations of $\psi$ are amplified, 
during inflation, with a scale-invariant spectrum.
Therefore, denoting by $\chi_{\psi}$ the fluctuation of $\psi$, we do 
know that the power spectrum will be $\delta_{\chi_{\psi}} \sim H_{\rm e}/2\pi$. 
If 
\begin{equation}
\frac{H_{\rm e}}{M_{\rm P}} < 2 \pi \frac{\psi_{\rm e}}{M_{\rm P}},
\label{con4}
\end{equation}
then the fluctuations of $\psi$ will be predominantly Gaussian. In the opposite 
case they will have some non-negligible non-Gaussian component. 
In order to have a nearly scale-invariant spectrum 
for  $\chi_{\psi}$, $m$ should always be 
 smaller than the curvature scale $H_{\rm e}$.
More precisely, looking at the evolution equation for 
$\chi_{\psi}$ (in conformal time) we are led to require  
\begin{equation}
m <\sqrt{2} H_{\rm e}.
\label{con3}
\end{equation}
Equation (\ref{con3}) is illustrated in Fig.\ref{constraints} 
with the dotted horizontal line. Finally, since the decay 
of $\psi$ should occur prior to BBN, an absolute 
lower bound on $m$, i.e. $m > 10 $ TeV, should be imposed.
Therefore, already from these considerations it is 
possible to say that the dashed area in Fig. \ref{constraints} 
is allowed. Thus,  the quintessence field has to 
become subdominant before the decay of $\psi$ if we want 
$\chi_{\psi}$ to be Gaussian with nearly scale-invariant spectrum.

\renewcommand{\theequation}{3.\arabic{equation}}
\setcounter{equation}{0}
\section{Quintessential evolution}
In order to satisfy the physical constraints 
for the consistency of the scenario, it is reasonable 
to require  that 
the decay of $\psi$  occur when $\varphi$ is already sub dominant. 
If this is the situation, $ H_{\rm d} > H_{\rm r}$ and, consequently,
\begin{equation}
\psi_{\rm e } ~\gaq \frac{1}{\sqrt{5} } m.
\end{equation}
It is important, for the purposes of the present investigation, 
to analyse in some detail,
the evolution of $\varphi$ around the curvature scales 
$H_{\rm d}$ and $H_{\rm r}$. 

After $H_{\rm d}$, the evolution of $\varphi$ results from
the interplay between the smallness of its potential term $V(\varphi)$ and 
the coherent oscillations of $\psi$, which tend to make the average expansion
matter-dominated.
The effect of the potential is, however, to 
introduce a growing mode in the evolution of $\varphi$.
This growing mode can be estimated by solving Eq. (\ref{phb}) 
with the inverse power-law potential of Eq. (\ref{potph}).
The solution is 
\begin{equation}
\varphi_{\rm g}(t) \simeq 3^{1/3} ~\lambda^{1/6} M^{4/3} t^{1/3},~~~~~ t_{\rm d} < t < t_{\rm r}.
\label{grow}
\end{equation}
For $t> t_{\rm r}$, when the Universe is effectively 
radiation-dominated, a similar solution holds: 
\begin{equation} 
\varphi_{\rm g}(t) \simeq 
\biggl(\frac{72}{5}\biggr)^{1/6} ~\lambda^{1/6} M^{4/3} t^{1/3},~~~~~ t > t_{\rm r}.
\end{equation}

The effective evolution of $\varphi$ depends upon the balance 
of the growing mode with the decaying modes obtained from 
Eq. (\ref{phb}) when the potential is neglected.
During  the regime where $\psi$ dominates and in  the subsequent 
radiation-dominated regime, the evolution of 	
$\varphi$ is given, respectively, by 
\begin{eqnarray}
&&\varphi_{\rm d}(t) \simeq \frac{\sqrt{2}~M_{\rm P}}{3} \biggl\{ \biggl[ 
\ln{\biggl(\frac{t_{\rm d}}{t_{\rm e}}\biggr)} + 1\biggr] - \frac{t_{\rm d}}{t}\biggr\},~~~~
t_{\rm d} < t < t_{\rm r},
\label{g1}\\
&& \varphi_{\rm r}(t) \simeq \frac{\sqrt{2}~M_{\rm P}}{3} \biggl\{ \biggl[ 
\ln{\biggl(\frac{t_{\rm d}}{t_{\rm e}}\biggr)} + 1 + \frac{t_{\rm d}}{t_{\rm r}}\biggr] 
-2 \frac{t_{\rm d}}{t_{\rm r}}\sqrt{\frac{t_{\rm r}}{t}}\biggr\},~~~~~~~~~
t > t_{\rm r}.
\label{g2}
\end{eqnarray}
Equations (\ref{g1}) and(\ref{g2}) are continuous in $t_{\rm r}$. Furthermore 
Eqs. (\ref{vfkin}) and (\ref{g1}) are continuous in $t_{\rm d}$.
As obtained in \cite{pv1} the growing solution never comes to dominate against the decreasing mode. 
If 
\begin{equation}
\lambda^{-1/2} \ln^3{\biggl(\frac{t_{\rm d}}{t_{\rm e}}\biggr)} \biggl(\frac{M_{\rm P}}{M}\biggr)^4 > 
\biggl(\frac{H_{0}}{M_{\rm P}}\biggr)^{-1},
\label{exp}
\end{equation}
the growing mode becomes dominant only {\em after} the present expansion time $t_{0} \sim H_{0}^{-1}$.
We will see that this condition is always verified because of the 
smallness of $\lambda$.

After $t_{\rm r}$ the quintessence field is nearly constant 
and its potential energy is given by 
\begin{equation}
V(\varphi_{\rm r}) \simeq \lambda \frac{M^8}{M_{\rm P}^4} 
\biggl[\ln{\biggl(\frac{t_{\rm d}}{t_{\rm e}}\biggr)}\biggr]^{-4}.
\end{equation}
The condition that  the potential energy in the quintessence 
field is comparable with the present energy density,
\begin{equation}
V(\varphi_{\rm r}) \simeq H_{0}^2 M_{\rm P}^2,
\end{equation}
fixes a relation among $M$ and  $\lambda$
\begin{equation}
\frac{M}{M_{\rm P}} \simeq \biggl(\frac{9}{2}\biggr)^{-1/4} 
\biggl(\frac{H_0}{M_{\rm P}}\biggr)^{1/4} \lambda^{-1/8} 
\biggl\{\ln{\biggl[ \frac{\sqrt{\lambda}}{5} \biggl(\frac{M_{\rm P}}{m}\biggr) 
\biggl(\frac{M_{\rm P}}{\psi_{\rm e}}\biggr)^2\biggr]}\biggr\}^{1/2}.
\end{equation}
Note that, for the typical parameters discussed so far, $M$ is always greater 
than $10^{5}$ GeV.

\renewcommand{\theequation}{4.\arabic{equation}}
\setcounter{equation}{0}
\section{Large scale fluctuations}
Having discussed the main features of the post-inflationary evolution,
it is now mandatory to understand the behaviour of the fluctuations.
Available calculations on the curvaton dynamics in different models
\cite{es,lw,us,sl,mor,bas,gor}
always deal with a post-inflationary phase, which is 
dominated by radiation. Here, as discussed in the previous sections, the scenario 
is different. Right after inflation the background
evolution is not determined by radiation but by the dynamics of $\varphi$,
whose potential is now very steep. The fluctuation 
of the metric induced by the fluctuations of $\varphi$ will be, at the 
onset of the post-inflationary epoch, very small since 
$H_{\rm e} < 10^{-6} M_{\rm P}$.  Hence, had we to classify the metric 
fluctuations at $t_{\rm e}$, we would say that their modes are of  the isocurvature type.
However, as time goes by, the curvature fluctuations, initially negligible at $t_{\rm e}$,
 will be driven to a constant value so that, much later, the initial 
isocurvature mode turns into an adiabatic one.

In order to describe this picture quantitatively,  we are led to consider 
the coupled system formed by the fluctuations of the metric and by 
the fluctuations of $\varphi$ and $\psi$ .
The fluctuations will  then be discussed
in the longitudinal gauge, where it  is particularly simple to 
relate the gauge-dependent quantities to gauge-invariant observables \cite{bran,bar}.

In the longitudinal gauge, the non-vanishing entries 
of the perturbed metric are 
\begin{equation}
\delta g_{00} = 2 \phi,~~~~~\delta g_{i j} = -2 a^2 \phi,
\end{equation}
where we used the fact that the fluctuations of the energy-momentum 
tensor are free of shear.
For the fluctuations of $\varphi$ and $\psi$, 
in the longitudinal gauge the following notation 
will also be adopted:
\begin{eqnarray}
&&\varphi \to \varphi + \delta\varphi ,~~~ \delta\varphi = \chi_{\varphi},
\nonumber\\
&& \psi \to \psi + \delta \psi, ~~~~~\delta\psi = \chi_{\psi}. 
\end{eqnarray}
With these notations the evolution equations of the 
fluctuations can be written as 
\begin{eqnarray}
&& \ddot{\phi} + 4 H \dot{\phi} + ( 2 \dot{H} + 3 H^2)  \phi = 
- \frac{3}{2 M_{\rm P}^2} \biggl[ (\dot{\varphi}^2 + \dot{\psi})^2 \phi
 - ( \dot{\psi} \dot{\chi}_{\psi}  + \dot{\varphi} \dot{\chi}_{\varphi}  )
+ \frac{\partial V}{\partial\varphi} \chi_{\varphi}
+ \frac{\partial W}{\partial \psi}  \chi_{\psi}\biggr],
\label{ij}\\
&& 3 H ( H \phi + \dot{\phi})- \frac{1}{a^2}\nabla^2\phi = -  \frac{3}{2 M_{\rm P}^2}
 \biggl[ -(\dot{\varphi}^2 + \dot{\psi})^2 \phi
 + ( \dot{\psi} \dot{\chi}_{\psi}  + \dot{\varphi} \dot{\chi}_{\varphi}  )
+ \frac{\partial V}{\partial\varphi} \chi_{\varphi}
+ \frac{\partial W}{\partial \psi}  \chi_{\psi}\biggr],
\label{oo}\\
&& \ddot{\chi}_{\varphi} + 3 H \dot{\chi}_{\varphi} - 
\frac{1}{a^2} \nabla^2 \chi_{\varphi} 
+ \frac{\partial^2 V}{\partial\varphi^2}  \chi_{\varphi} -
 4 \dot{\varphi}\dot{\phi} + 2 \frac{\partial V}{\partial \varphi}  \phi =0,
\label{chph}\\
&& \ddot{\chi}_{\psi} + 3 H \dot{\chi}_{\psi} - 
\frac{1}{a^2} \nabla^2 \chi_{\psi} 
+ \frac{\partial^2 W}{\partial\psi^2} \chi_{\psi} -
 4 \dot{\psi}\dot{\phi} + 2 \frac{\partial W}{\partial \psi}  \phi=0,
\label{chps}
\end{eqnarray}
 where (\ref{ij}) and (\ref{oo}), come, 
respectively, from the $(i,j)$ and $(0,0)$ 
components of the perturbed Einstein equations and (\ref{chph}) and (\ref{chps}) 
describe the evolution of the  inhomogeneities in the inflaton/quintessence 
field and in the curvaton field.
Equations (\ref{ij})--(\ref{chps}) are subjected to the momentum constraint
\begin{equation}
H \phi +\dot{\phi} = \frac{3}{2 M_{\rm P}^2} ( \dot{\varphi} \chi_{\varphi} + 
\dot{\psi} \chi_{\psi} ).
\label{0i}
\end{equation}
Using Eq. (\ref{0i}) together with Eqs. (\ref{ij})--(\ref{chps}), it is possible 
to obtain a nicer form of the perturbation equations \cite{hw1,hw2,hw3,tn}:
\begin{eqnarray}
&&\ddot{v}_{\varphi} + 3 H \dot{v}_{\varphi} - \frac{1}{a^2} \nabla^2 v_{\varphi} +
\biggl[ \frac{\partial^2 V}{\partial\varphi^2} - \frac{3}{M_{\rm P}^2 a^3}\frac{\partial}{\partial t} 
\biggl( \frac{a^3}{H} \dot{\varphi}^2\biggr)\biggr] v_{\varphi} 
- \frac{3}{M_{\rm P}^2 a^3}\frac{\partial}{\partial t} 
\biggl( \frac{a^3}{H} \dot{\varphi} \dot{\psi}\biggr) v_{\psi}=0,
\label{vph}\\
&&\ddot{v}_{\psi} + 3 H \dot{v}_{\psi} 
- \frac{1}{a^2} \nabla^2 v_{\psi} +
\biggl[ \frac{\partial^2 W}{\partial\psi^2} 
- \frac{3}{M_{\rm P}^2 a^3}\frac{\partial}{\partial t} 
\biggl( \frac{a^3}{H} \dot{\psi}^2\biggr)\biggr] v_{\psi} 
- \frac{3}{M_{\rm P}^2 a^3}\frac{\partial}{\partial t} 
\biggl( \frac{a^3}{H} \dot{\varphi} \dot{\psi}\biggr) v_{\varphi}=0,
\label{vps}
\end{eqnarray}
where 
\begin{eqnarray}
&& v_{\varphi} = \chi_{\varphi} + \frac{\dot{\varphi}}{H} \phi,
\label{defvph}\\
&& v_{\psi} = \chi_{\psi} + \frac{\dot{\psi}}{H} \phi.
\label{defvps}
\end{eqnarray}
Eqs. (\ref{vph}) and (\ref{vps}) can be generalized to the case of 
an arbitrary number of fields \cite{hw1,hw2}: in this case the number of equations will
clearly match the number of fields but the relative structure of the equations will be the same.

Eqs. (\ref{oo})--(\ref{chps}) or, equivalently, 
Eqs. (\ref{vph}) and(\ref{vps}), have to be  studied and solved  along the 
different stages of the evolution of the background. 
In terms of $v_{\varphi}$ and $v_{\psi}$, the spatial 
curvature perturbation can be written as 
\begin{equation}
\zeta = - \frac{H}{\dot{\varphi}^2 + \dot{\psi}^2} 
\bigl[ \dot{\varphi} v_{\varphi}  +\dot{\psi} v_{\psi} \bigr].
\label{zetdef1}
\end{equation}
The variable $\zeta$ is related also to  $\phi$ by the usual expression
\begin{equation}
\zeta=  \frac{H}{\dot{H}}(H \phi + \dot{\phi}) -\phi.
\label{zetord}
\end{equation}
Eq. (\ref{zetord}) can be obtained from Eq. (\ref{zetdef1}) (or vice versa) 
by using the momentum constraint (\ref{0i}) expressed in terms of  the 
of $v_{\varphi}$ and $v_{\psi}$ given in Eqs. (\ref{defvph}) and (\ref{defvps}).

An interesting (complementary)
 strategy in order to solve the system (\ref{ij})--(\ref{chps}) and 
(\ref{0i}) is to write down directly the evolution equation for $\zeta$.
Mutiplying Eq. (\ref{oo}) by the sound of speed and subtracting it from  
 Eq. (\ref{ij}) we obtain, at large scales \cite{ks},
\begin{equation}
\frac{d \zeta}{d t} = - \frac{H}{\dot{\psi}^2 + \dot{\varphi}^2} 
\delta p_{\rm nad},
\label{zeq}
\end{equation}
where $\delta p_{\rm nad}$ is, in our case,
\begin{equation}
\delta p_{\rm nad} = ( c_{s}^2 -1) \phi ( \dot{\varphi}+ \dot{\psi}) + 
( 1 - c_{s}^2) ( \dot{\varphi} \dot{\chi}_{\varphi} + \dot{\psi} \dot{\chi}_{\psi}) 
- ( 1 + c_{s}^2) \biggl( \frac{ \partial V}{\partial\varphi} \chi_{\varphi} + 
\frac{\partial W}{\partial\psi} \chi_{\psi}\biggr),
\label{dp1}
\end{equation}
with 
\begin{equation}
c_{s}^2 = \frac{\dot{p}}{\dot{\rho}}=
1 + \frac{2}{3 H( \dot{\varphi}^2 + \dot{\psi}^2)}\biggl( \frac{\partial V}{\partial\varphi} 
\dot{\varphi} + \frac{\partial W}{\partial \psi}\dot{\psi} \biggr).
\label{cs1}
\end{equation}
In the first equality of Eq. (\ref{cs1}) $p$ and $\rho$ are, respectively, the total 
pressure and energy density of the system written in terms of the two background fields $\varphi$ and $\psi$.
Notice that in order to get to (\ref{cs1}) the background equations of motion (\ref{b1})--(\ref{psb}) have been used.

The initial conditions of the system (\ref{oo})--(\ref{chps}) after the end of inflation 
are dictated by the smallness of $\lambda$. Going to Fourier space and 
considering only the super-horizon scales, the initial 
conditions of the system are, for $\lambda \ll 10^{-14}$, 
\begin{equation}
\phi(k,t_{\rm e}) =0, ~~~~\chi_{\varphi}(k,t_{\rm e}) =0,~~~~~
\chi_{\psi}(k,t_{\rm e})= \frac{H_{\rm e}}{2 \pi}.
\label{incon0}
\end{equation}
In terms of $v_{\varphi}$ and $v_{\psi}$,   we have 
at the end of inflation, from Eqs. (\ref{vph}) and (\ref{vps}), on super-horizon scales,
\begin{equation}
v_{\varphi}(k,t_{\rm e}) =0 ,~~~~~~
v_{\psi}(k,t_{\rm e}) =\chi_{\psi}(k,t_{\rm e}), 
\label{incon1}
\end{equation}
as  is made clear by inserting Eqs. (\ref{incon0}) into  Eqs. (\ref{vph}) and (\ref{vps}).
Physically Eqs. (\ref{incon0}) and (\ref{incon1}) guarantee 
the absence of adiabatic modes at $t_{\rm e}$. This aspect can also be 
appreciated by looking at the initial conditions for $\zeta$
\begin{equation}
\zeta(k,t_{\rm e}) =0,
\label{incon2}
\end{equation} 
as they follow from Eq. (\ref{zetdef1}).

\renewcommand{\theequation}{5.\arabic{equation}}
\setcounter{equation}{0}
\section{Evolution during the kinetic phase} 
Using the initial conditions given by Eqs. (\ref{incon0})--(\ref{incon2}), 
the evolution of the fluctuations can be solved in slightly different but, ultimately, 
equivalent ways. In the following, as a first step,  the asymptotes for the 
evolution of the fluctuations will be discussed analytically in the  vicinity of $t_{\rm e}$. 
The consistency of the analytical solutions with the results of the numerical integration 
is an important check to be done. Following recent techniques developed 
in a different framework \cite{us}, the solutions in 
the vicinity of $t_{\rm e}$ can be obtained without 
specifying the form of the potential
for the field $\psi$. Later on, in order to perform the numerical
integration, the case of massive potential will be mainly discussed.

\subsection{The initial stages around $t_{\rm e}$ for generic potential}

During this phase the field $\psi$ slowly rolls down its own potential
$W(\psi)$, and the solution to Eq. (\ref{psb}) is given by 
(\ref{psap}). The same expansion in the gradients 
of the potential $W(\psi)$ can be used in order to get the approximate 
evolution of the fluctuations.

Eqs. (\ref{vph}) and (\ref{vps}) can then be approximately 
solved by keeping the leading terms in the derivatives of 
$W(\psi)$. As done in the case of the background evolution, an equation 
analogous to  (\ref{psap}) can also be obtained for the canonical 
perturbation variable $v_{\psi}$.
Eq. (\ref{vps}) can then be expanded in gradients of the potential leading 
to the following approximate equation
\begin{equation}
\dot{v}_{\psi} = - \frac{1}{6 H} \frac{\partial^2 W}{\partial \psi^2} v_{\psi},
\label{vps3}
\end{equation}
whose solution implies that $v_{\psi}$ is approximately constant. 
For instance, inserting into Eq. (\ref{vps3}) the explicit 
expression for $W(\psi)$ given in Eq. (\ref{potps}), we get 
\begin{equation}
v_{\psi}(k,t) \simeq v_{\psi}(k,t_{\rm e}) \biggl[1 - \frac{m^2}{4}(t^2 -t_{\rm e}^2)\biggr], ~~~~t<t_{\rm m}.
\label{approxvps}
\end{equation}
This expression has been verified numerically for various values of the physical parameters.
Using Eq. (\ref{approxvps}),  Eq. (\ref{vph})  can be solved in the same approximation.
Note, in fact, that in Eq. (\ref{vph}) 
the term in square brackets is subleading with respect to the 
other terms for two separate reasons. First $V_{,\varphi\varphi}$ is negligible 
in its own right, given the smallness of $\lambda$ and recalling that $ M/M_{\rm P}$ is 
${\cal O}(10^{-13})$. Second, since   $a^{3} \dot{\varphi}^2/H $  is constant,
its time derivative appearing in  Eq. (\ref{vph}) is also negligible.
The approximate evolution of $v_{\varphi}(k,t)$ is then given, at large scales, by
\begin{equation}
\ddot{v}_{\varphi} + 3 H \dot{v}_{\varphi} 
- \frac{3}{M_{\rm P}^2 a^3}\frac{\partial}{\partial t} 
\biggl( \frac{a^3}{H} \dot{\varphi} \dot{\psi}\biggr) v_{\psi}=0,
\label{vph2}
\end{equation}
whose solution, using Eqs. (\ref{psap}) and  (\ref{vps3}), becomes 
\begin{equation}
v_{\varphi}(k,t) \simeq - \frac{3}{4 M_{\rm P}^2} \biggl(\frac{\dot{\varphi}}{H}\biggr) 
\frac{\partial W}{\partial\psi} v_{\psi}(k,t)
[a^6(t) -  a^{6}(t_{\rm e})].
\label{solvph}
\end{equation}
Given that during the kinetic phase $\dot{\varphi}\simeq H$, then $v_{\varphi}(k,t) \propto a^{6}$.
Again, in the case of Eq. (\ref{potps}), from Eq. (\ref{solvph})  we obtain
\begin{equation}
v_{\varphi}(k,t) \simeq -\frac{3 \sqrt{2}}{4} \biggl(\frac{\psi_{\rm e}}{M_{\rm P}}\biggr) m^2( t^2 - t_{\rm e}^2) 
\chi_{\psi}(k,t_{\rm e}),
\label{solvph1}
\end{equation} 
where, according to Eq. (\ref{incon1}),  $v_{\psi}(k,t_{\rm e}) = \chi_{\psi}(k,t_{\rm e})$ has been used.
With the results of Eqs. (\ref{approxvps}) and (\ref{solvph1}),  
from Eqs. (\ref{defvph})--(\ref{zetdef1}) we can also obtain the approximate 
form of the evolution of $\phi_{k}(t)$, $\zeta_{k}(t)$ and $\chi_{\varphi}(k,t)$:
\begin{eqnarray}
&& \zeta_{k}(t) \simeq \frac{3}{2 M_{\rm P}^2} \frac{\partial W}{\partial\psi} 
\chi_{\psi}(k,t_{\rm e}) [a^{6}(t) -  a^{6}(t_{\rm e})],
\label{zet1}\\
&& \phi_{k}(t) \simeq - \frac{3}{10} \zeta_{k}(t), 
\label{phi1}\\
&& \chi_{\varphi}(k,t) \simeq - \frac{3}{10 M_{\rm P}^2} \biggl(\frac{\dot{\varphi}}{H}\biggr) 
\frac{\partial W}{\partial\psi}\chi_{\psi} [a^6(t) -a^6(t_{\rm e})].
\label{chph1}
\end{eqnarray}

The same results derived in Eqs. (\ref{zet1})--(\ref{chph1}) and based on the solution 
of Eqs. (\ref{vph}) and (\ref{vps}),
 with the initial conditions dictated by Eq. (\ref{incon1}), 
 can be obtained by integrating Eqs. (\ref{oo})--(\ref{chps}). As a check of the consistency 
of the approach, it is in fact useful to insert Eqs. (\ref{phi1}) and (\ref{chph1}) back 
into Eqs. (\ref{ij})--(\ref{chps}) and see that they are satisfied during the kinetic phase 
and prior to the oscillations of $\psi$. 
Furthermore, the evolution of $\zeta_{k}$, as obtained 
in Eq. (\ref{zet1}), can be also obtained directly from Eq. (\ref{zeq}).
In fact, during the kinetic phase $ \dot{\psi} \ll \dot{\varphi}$ and, 
from Eq. (\ref{cs1}) evaluated in the kinetic limit:
\begin{equation}
c_{s}^2 \simeq 1 - \frac{1}{18 M_{\rm P}^2 H^4} \frac{\partial W}{\partial\psi}. 
\end{equation}
Hence, from Eq. (\ref{dp1}) we find
\begin{equation}
\delta p_{\rm nad} = - 2 \frac{\partial W}{\partial \psi} \chi_{\psi}.
\label{dpkin}
\end{equation}
Inserting (\ref{dpkin}) into Eq. (\ref{zeq}) and performing the integral,
we get exactly Eq. (\ref{zet1}).
The analytic expressions derived so far allow a full control of the initial 
conditions of the system in the vicinity of $t_{\rm e}$.

\subsection{Evolution for $t_{\rm e} < t < t_{\rm d}$}

\begin{figure}
\begin{center}
\begin{tabular}{|c|c|}
      \hline
      \hbox{\epsfxsize = 7.5 cm  \epsffile{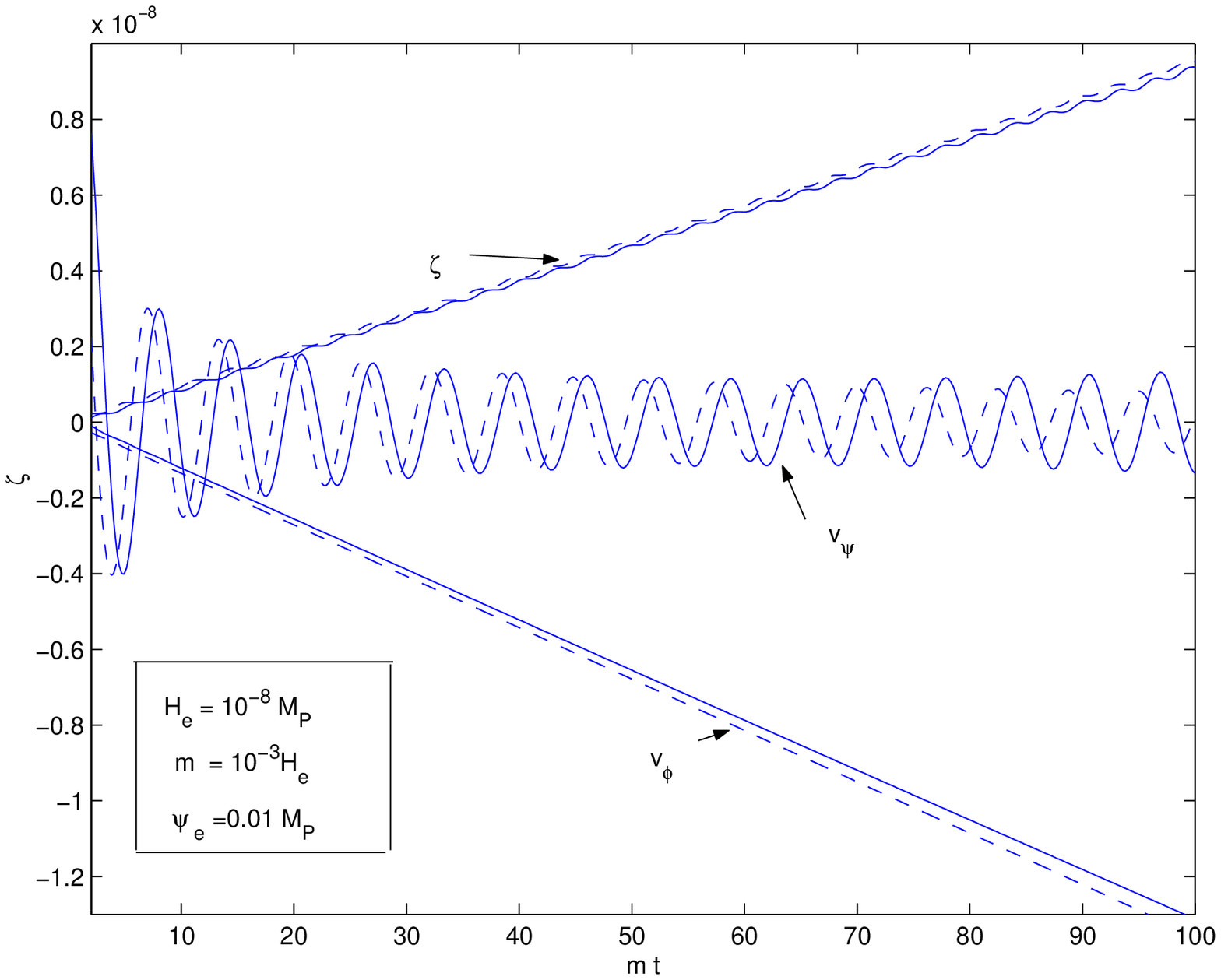}} &
      \hbox{\epsfxsize = 7.5 cm  \epsffile{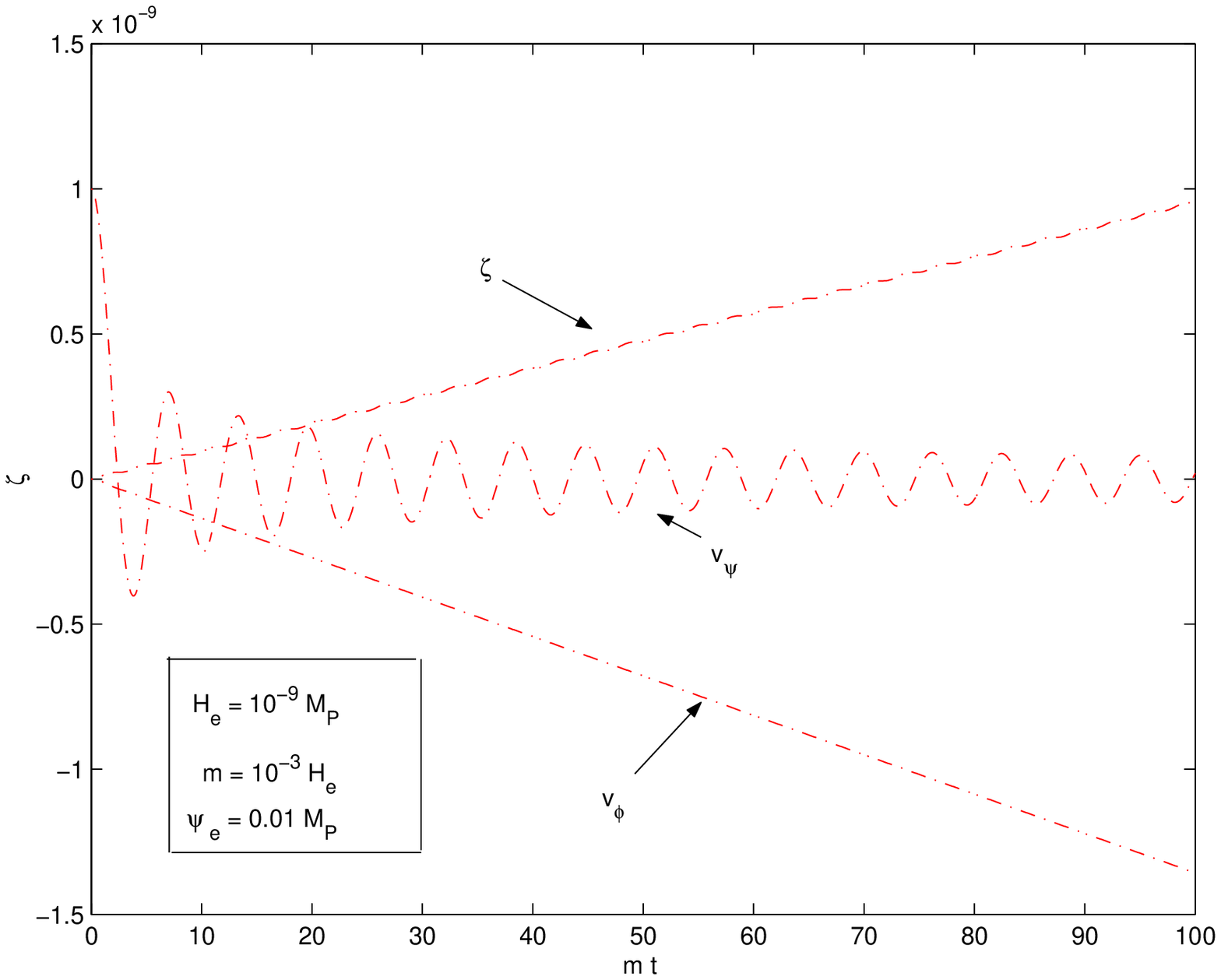}}\\
      \hline
\end{tabular}
\end{center}
\caption[a]{The result of the numerical integration for the evolution of 
the fluctuations are illustrated for the case $M/M_{\rm P} = 10^{-13}$ and 
for a set of fiducial parameters chosen within the shaded region of Fig.
 \ref{constraints}. In the left plot the analytical results (dashed lines) 
obtained in Eqs. (\ref{solvps}), (\ref{exactvph}) and (\ref{zetfina1}) 
are compared with the numerical ones (full lines).}
\label{f2}
\end{figure}

After the onset of the kinetic phase at $t_{\rm e}$, but 
before the dominance of $\psi$ at $t_{\rm d}$, the evolution of 
the system can be solved numerically 
for various sets of initial conditions; an example of this behaviour 
is reported in Fig. \ref{f2} for the case of the potential given in Eq. (\ref{potps}).

It will now be shown that the numerical 
evolution can be very accurately reproduced analytically by direct integration 
of the evolution equation in a well defined approximation 
scheme. The evolution of the fluctuations will be analysed first 
using Eqs. (\ref{vph}) and (\ref{vps}) and then using directly 
the evolution equation for $\zeta$, i.e. Eq. (\ref{zeq}).

In the case of the potential (\ref{potps}), 
Eqs. (\ref{vph}) and (\ref{vps}) can be written as 
\begin{eqnarray}
&&\frac{d}{dt} \biggl( a^3 \dot{v}_{\varphi} \biggr) = - \frac{3\sqrt{2}}{M_{\rm P}} m^2 \psi v_{\psi},
\label{vph1a}\\
&& \ddot{v}_{\psi} + 3 H \dot{v}_{\psi} + m^2 \biggl[ 1 + \frac{6}{H M_{\rm P}^2} \dot{\psi} \psi\biggr] v_{\psi} 
+ \frac{3 \sqrt{2}}{M_{\rm P}} m^2 \psi v_{\varphi}=0,
\label{vps1a}
\end{eqnarray}
where Eqs. (\ref{vfkin}) and (\ref{psb})  have been used.
Recalling now the exact solution for the evolution of (\ref{psb}), i.e. Eq. (\ref{ansolpsi}), 
it can be easily checked that, in Eq. (\ref{vps1a}):
\begin{equation}
\frac{6}{H M_{\rm P}^2} \dot{\psi} \psi \sim \frac{18 t}{M_{P}^2} \psi_{\rm e}^2 m < 1
\end{equation}
for $t < t_{\rm d}\sim m^{-1} (M_{\rm P}/\psi_{\rm e})^2$.
Neglecting the term containing $v_{\varphi}$ in Eq. (\ref{vps1a}), the 
solution for the evolution of $v_{\psi}$ is given by 
\begin{equation}
v_{\psi}(k,t) = \frac{v_{\psi}(k,t_{\rm e})}{J_{0}( x_{\rm e})} J_{0}(x),
\label{solvps}
\end{equation}
where $x = mt$ and $J_{0}(x)$ is the Bessel function of order 0 \cite{magn}. 
Eq. (\ref{solvps}) reproduces exactly 
the numerical solutions for $v_{\psi}$ reported, for  a particular case,
in Fig. \ref{f2}. Since it is always true that
$m t_{\rm e} \ll 1$  for the constraints displayed in Fig. \ref{constraints}, 
the interesting limits of Eq. (\ref{solvps}) are for 
$m t \ll 1$ and $m t \gg 1$. In the limit $m t \ll 1$ 
 Eq. (\ref{solvps}) reproduces, as it should, the 
time dependence obtained in 
Eq. (\ref{approxvps}). For 
$m t \gg 1$ we have \cite{magn}, from Eq. (\ref{solvps}):
\begin{equation}
v_{\psi}(k,t) = v_{\psi}(k,t_{\rm e}) \sqrt{\frac{2}{\pi x}} \cos{\biggl(x - \frac{\pi}{4}\biggr)}
\label{appoxsol2}
\end{equation}
Note the similarity between Eqs. (\ref{solvps}) and (\ref{ansolpsi}), which is a simple
consequence of the quadratic form of the potential.

Inserting now Eq. (\ref{solvps}) in Eq. (\ref{vph1a}), and integrating a first 
time between $t_{\rm e}$ and a generic time $t$ we get 
\begin{equation}
\frac{d v_{\varphi}}{d x} = - \frac{3 \sqrt{2}}{2 M_{\rm P}} 
\frac{\psi_{\rm e} v_{\psi}(k,t_{\rm e})}{J_{0}( x_{\rm e})}\biggl[ x \biggl(J_{0}^2(x) + J_{1}^2(x)\biggr) 
- \frac{F(x_{\rm e})}{x}\biggr],
\label{firstint}
\end{equation}
where $F(x_{\rm e}) =x_{\rm e}^2 [ J_{0}(x_{\rm e})^2 + J_{1}(x_{\rm e})^2]$.

Direct integration of Eq. (\ref{firstint}) implies that 
\begin{equation}
v_{\varphi}(k, x) = - \frac{ 3 \sqrt{2}}{4} \frac{\psi_{\rm e} v_{\psi}(k,t_{\rm e}) }{J_{0}(x_{\rm e})^2}
\biggl[ x^2 \biggl( J_{0}^2(x) + 2 J_{1}^2(x) - J_{0}(x) J_{2}(x) \biggr) - G(x_{\rm e}) - 2 F(x_{\rm e})\biggr],
\label{exactvph}
\end{equation}
where $G(x_{\rm e}) = x_{\rm e}^2 [ J_{0}^2(x_{\rm e}) + 2 J_{1}^2(x_{\rm e}) - J_{0}(x_{\rm e}) J_{2}(x_{\rm e}) ]$
and where, as usual, $x_{\rm e} \sim m/H_{\rm e}$.
Taking the limit for large $x$  of Eq. (\ref{exactvph}) the following result can be obtained 
\begin{equation}
v_{\varphi}(k, t) = - \frac{6 \sqrt{2}}{\pi} \psi_{\rm e} v_{\psi}(k,t_{\rm e}) (m t) + \frac{3\sqrt{2}}{4}  
\psi_{\rm e} v_{\psi}(k,t_{\rm e}) (m t_{\rm e})^2 \ln{\frac{t}{t_{\rm e}}},
\end{equation}
showing off the linear growth of $v_{\varphi}$ with time.

Inserting now the solutions for $v_{\psi}$ and $v_{\varphi}$ given, respectively, in Eqs. (\ref{solvps}) and (\ref{exactvph}) 
into Eq. (\ref{zetdef1}), the evolution of $\zeta$ for $t< t_{\rm d}$, turns out to be, for $m t > 1$,
\begin{equation}
\zeta(k,t) \simeq \frac{6}{\pi} \biggl(\frac{\psi_{\rm e}}{M_{\rm P}}\biggr)
 \biggl[\frac{v_{\psi}(k, t_{\rm e})}{M_{\rm P}}\biggr] (m t ).
\label{zetfina1}
\end{equation}
Note that, for $t= t_{\rm d}$, Eq. (\ref{zetfina1}) 
gives 
\begin{equation}
\zeta(k,t_{\rm d}) = \frac{2}{5 \pi} \biggl(\frac{v_{\psi}(k, t_{\rm e})}{\psi_{\rm e}}\biggr).
\end{equation}
Using the result of Eq. (\ref{zetfina1}) into Eq. (\ref{zetord})
the evolution for the large-scale modes of the metric fluctuations can be obtained:
\begin{equation}
\phi(k,t) \simeq - \frac{18}{7 \pi}\biggl(\frac{\psi_{\rm e}}{M_{\rm P}}\biggr) 
\biggl[\frac{v_{\psi}(k, t_{\rm e})}{M_{\rm P}}\biggr] (m t ).
\end{equation}
Again we verified that 
 the same results can be obtained directly by integrating the Hamiltonian constraint (\ref{oo}).

The time has come to compare the accuracy of the analytical expressions 
derived in Eqs. (\ref{solvps}), (\ref{exactvph}) and (\ref{zetfina1}). 
These expressions have been plotted in Fig. \ref{f2} (left plot) with the dashed lines.
For the same set of parameters (and with the full line) the outcome of the numerical
integration has been reported. In the right plot the results for a different set of
parameters are displayed.
The analytical expressions match rather accurately the numerical results.
Notice that the small wiggles in the evolution of $\zeta(k, t)$, modulating the linear growth, 
are due to the fact that 
we decided to plot the full expression of $\zeta(k,t)$, which contains Bessel functions, 
and not only its asymptotic limit for $m t >1$.
The  same evolution for $\zeta(k,t)$ derived in this section 
can be directly inferred from Eq. (\ref{zeq}), recalling 
the approximate form of $\delta p_{\rm nad}$. This 
calculation is reported in the appendix. 

It is now appropriate to compare the situation of low-scale quintessential 
inflation with the situation occurring in the more conventional case 
of curvaton models where the inflationary phase is suddenly followed 
by the radiation-dominated phase. 
In this case the $\zeta(k,t)$ variable grows as $\sqrt{t}$ (i.e. linearly in conformal time)
 before the curvaton becomes dominant.
Here we found that the growth is linear in {\em cosmic} time. In spite 
of this difference, the final amplitude of $\zeta(k,t)$ is given approximately 
by $\chi_{\psi}(k,t_{\rm d})/\psi_{\rm e}$. In fact,
 in the case of a quadratic potential 
evolving in a radiation-dominated environment, 
\begin{equation}
\dot{\zeta} \sim \biggl(\frac{\psi_{\rm e}}{M_{\rm P}}\biggr) 
\biggl[\frac{\chi_{\psi}(k,t_{\rm e}}{M_{\rm P}} \biggr] H a(t),
\end{equation}
where we now have $a(t) \sim \sqrt{t}$. Integrating once the previous formula we get 
$\zeta(k,t) \sim \sqrt{t}$. If the evolution occurs during radiation, the 
 curvaton will become dominant 
at a typical curvature scale $H_{\rm d} \sim m (\psi_{\rm e}/M_{\rm P})^4$ \cite{us}.
Using this result, the amplitude of $\zeta(k,t_{\rm d}) $ is given, as previously anticipated,
 by $\chi_{\psi}(k,t_{\rm d})/\psi_{\rm e}$.  

\renewcommand{\theequation}{6.\arabic{equation}}
\setcounter{equation}{0}
\section{Dominance of $\psi$} 

Since $\psi$ starts dominating the background at $t_{\rm d}$, 
for $t > t_{\rm d}$ the evolution of 
the system is  described 
by the following set of equations 
\begin{eqnarray}
&& H^2 M_{\rm P}^2 = \biggl[ \frac{\dot{\psi}^2}{2} + m^2 \psi^2\biggl],
\label{psdom1}\\
&& \dot{H} M_{\rm P}^2 = - \frac{3}{2} \dot{\psi}^2,
\label{psidom2}\\
&& \ddot{\psi} + 3 H \dot \psi + m^2 \psi =0.
\label{psdom3}
\end{eqnarray}
For large times Eqs. (\ref{psdom1})--(\ref{psdom3})
lead to an effectively matter-dominated phase where 
the oscillations of $\psi$,
\begin{equation}
\psi(t) \simeq \psi(t_{\rm d}) \frac{\cos{m t}}{ (H_{\rm d} t)^2},
\label{pc1}
\end{equation}
induce oscillations in the Hubble parameter and in the scale factor, 
which increases, on average, as $t^{2/3}$. 

It is not difficult to see that, under these conditions, the evolution 
of $v_{\psi}$ is dominated by a constant mode. In Eq. (\ref{vps}) 
the term containing $v_{\varphi}$ is always 
suppressed for large times since,from Eq. (\ref{g1})
\begin{equation}
\dot{\varphi} \simeq \frac{\sqrt{2}}{3} M_{\rm P} \frac{t_{\rm d}}{t^2},
\label{pc2}
\end{equation}
and the quintessence field goes very rapidly to a constant.
Thus, in this regime, the evolution of $v_{\psi}$ is given by 
\begin{equation}
v_{\psi}(k, t) = v_{\psi}(k,t_{\rm d}) \frac{\dot{\psi}}{H}.
\label{pc3}
\end{equation}
Eq. (\ref{vph}) allows us to deduce that
\begin{equation}
v_{\varphi}(k,t) \sim t^{-3}.
\label{pc4}
\end{equation} 
Inserting now Eqs. (\ref{pc1})--(\ref{pc4}) into (\ref{zetdef1}) 
we find that, for $t> t_{\rm d}$, $\zeta$ is frozen 
to its constant value.
Therefore, right before the decay of $\psi$, the metric fluctuation will be given 
by 
\begin{equation}
\phi(k,t_{\rm r}) \sim \frac{\chi_{\psi}(k, t_{\rm e})}{\psi_{\rm e}}.
\end{equation}

After $t_{\rm r}$ the field $\psi$ decays and the evolution of the 
generated adiabatic mode becomes standard, namely we have the 
fluctuations of the quintessence field evolving in a radiation-dominated environment
together with the constant mode of $\zeta$.
The evolution equations for the fluctuations of the quintessence field 
will then obey, at large scales,  
\begin{equation}
\ddot{\chi}_{\varphi} + 3 \dot{H} \dot{\chi}_{\varphi} + \frac{\partial ^2 V}{\partial\varphi^2} \chi_{\varphi} 
= 4 \dot{\varphi} \dot{\phi} - 2 \frac{\partial V}{\partial \varphi} \phi.
\label{qiev}
\end{equation}
The other equations are the standard ones, namely 
\begin{eqnarray}
&&-3 H ( H \phi + \dot{\phi}) = \frac{3}{2 M_{\rm P}^2}\rho_{\rm r} \delta_{\rm r} + 
\frac{3}{2 M_{\rm P}^2} \biggl[ - \phi \dot{\varphi}^2 + \dot{\varphi} \dot{\chi}_{\varphi} + 
\frac{\partial V}{\partial \varphi} \chi_{\varphi} \biggr],
\label{r1}\\
&& \ddot{\phi} + 4 H \dot{\phi} + ( 3 H^2 + 2 \dot{H}) \phi = \frac{1}{2 M_{\rm P}^2} \rho_{\rm r} \delta_{\rm r} - 
\frac{3}{2 M_{\rm P}^2} [ \dot{\varphi}^2 \phi - \dot{\chi}_{\varphi} + \frac{\partial V}{\partial\varphi} \chi_{\varphi}\biggr]
\label{r2}\\
&& H \phi + \dot{\phi} = \frac{3}{2 M_{\rm P}^2} \biggl[ \dot{\varphi} \chi_{\varphi} 
+ \frac{4}{3} \rho_{\rm r} u_{\rm r}\biggr],
\label{r3}\\
&& \dot{\delta}_{\rm r} - 4\dot{\phi} =0,
\label{r4}\\
&& \dot{u}_{\rm r} - \frac{1}{4} \delta_{\rm r} - \phi=0,
\label{r5}
\end{eqnarray}
where $\delta_{\rm r} = \delta \rho_{\rm r}/\rho_{\rm r}$ and $u_{\rm r}$ is the velocity potential.

As discussed in the context of the quintessential evolution of the background, the quintessence 
field, in the present model, does not lead to a tracking behaviour, as also noticed in 
 \cite{pv1}. Thus, the evolution 
of the fluctuations during the radiation-dominated stage of expansion will effectively be the one 
implied by a standard cosmological term. In fact we can recall, from Eq. (\ref{g2}), that 
$\varphi$ approaches a constant value as $t^{-1/2}$ while the potential 
is constant. Then, from Eq. (\ref{qiev}) it can be deduced that also $\chi_{\varphi} \sim t^{-1/2}$ at large scales.
As a consequence, combining Eqs. (\ref{r1}) and (\ref{r2}) we have, at large scales, 
\begin{equation}
\ddot{\phi} + 5 H \dot{\phi} + 2 (\dot{H} + 2 H^2) \phi =0.
\end{equation}
leading to the usual constant mode which was present prior to matter-radiation equality, a known 
feature of these types of models \cite{lv}.

\renewcommand{\theequation}{5.\arabic{equation}}
\setcounter{equation}{0}
\section{Concluding remarks} 
In the present paper, low-scale quintessential inflationary models have been analysed.
The set-up of the model is closely related to the one proposed by Peebles and Vilenkin. 
Even if the curvature scale at the end of inflation is much smaller than 
$10^{-6} ~M_{\rm P}$, curvature perturbations of correct amplitude can be 
generated. In these models, the 
quintessence and the inflaton field are identified. The consistency of the background evolution 
imposes a number of constraints on the various parameters. 
Assuming a specific form of the inflaton/quintessence potential and a massive 
curvaton, the evolution 
of the fluctuations has been followed  through the 
different stages of the model. Even if, for some region of the parameter space of the model, 
 non-Gaussian (adiabatic)
fluctuations can be generated, the case of the Gaussian adiabatic mode has been discussed. 
The curvaton evolution occurring in low-scale quintessential inflation
is different from the one possibly obtained in the case when the 
radiation-dominated evolution follows immediately inflation. In low-scale quintessential inflation, prior to curvaton
dominance, curvature perturbations grow linearly in cosmic time. On the contrary, if inflation is immediately followed by 
a radiation-dominated stage, then curvature fluctuations grow as $\sqrt{t}$.
  
In spite of the fact that low-scale quintessential inflationary models share essential analogies with 
the prediction of ordinary quintessential inflation there are also relevant differences between them. 
In ordinary quintessential inflationary models, the reheating is mainly gravitational. In the present case the 
reheating is triggered by the curvaton decay. In ordinary inflationary models 
a large background of gravitational waves should be expected in the GHz region \cite{mg1,mg2}. In the 
case of low-scale quintessential inflation, gravitational waves are negligible 
over all the frequencies of the spectrum.

\section*{Acknowledgments}
The author wishes to thank, V. Bozza, M. Gasperini and G. Veneziano for useful discussions.

\newpage
\begin{appendix}

\renewcommand{\theequation}{A.\arabic{equation}}
\setcounter{equation}{0}
\section{Evolution of $\zeta$ during the kinetic phase} 
In this appendix the analytical solution 
of Eq. (\ref{zeq}) will be obtained for $t_{\rm e}< t < t_{\rm d}$.
During the kinetic phase, $\dot{\varphi}^2 \gg \dot{\psi}^2$. Furthermore, 
as discussed in the paper
\begin{equation}
\delta p_{\rm nad} \simeq - 2 \frac{\partial W}{\partial\psi} \chi_{\psi} = 
- 2 m^2 \psi \chi_{\psi}.
\end{equation}
Thus, from Eq. (\ref{zeq}), we obtain that 
\begin{equation}
\frac{d \zeta}{d t} = \frac{2 H}{\dot{\varphi}^2} m^2 \psi \chi_{\psi}.
\label{zeqa}
\end{equation}
For  $t_{\rm e}< t < t_{\rm d}$, the approximate evolution 
of $\psi$ and $\chi_{\psi}$ is given by 
\begin{eqnarray}
&& \psi(t) = \psi_{\rm e} \frac{\sqrt{t}}{a^{3/2}} \frac{J_{0}(m t)}{J_0(m t_{\rm e})},
\nonumber\\
&& \chi_{\psi}(t) = \chi_{\psi}(k,t_{\rm e} )
\frac{\sqrt{t}}{a^{3/2}} \frac{J_{0}(m t)}{J_0(m t_{\rm e})}.
\end{eqnarray}
Inserting these solutions into Eq. (\ref{zeqa}) we get
\begin{equation}
\zeta(k,t) =  \frac{3}{2} \biggl(\frac{\psi_{\rm e}}{M_{\rm P}}\biggr) 
 \biggl(\frac{\chi_{\psi}(k,t_{\rm e})}{M_{\rm P}}\biggr) \biggl[ (m t)^2 \biggl(J_{0}^2(m t) + J_{1}^2(m t )\biggr) -
(m t_{\rm e})^2  \biggr].
\label{Eqzf}
\end{equation}
where we used the fact that $m t_{\rm e}\ll 1$ because of the constraint stemming from Eq. (\ref{con3}).
Taking the limit of Eq. (\ref{Eqzf}) for $ m t\gg 1$ we obtain 
\begin{equation}
\zeta(k,t) = \frac{6}{\pi} \biggl(\frac{\psi_{\rm e}}{M_{\rm P}}\biggr) 
 \biggl(\frac{\chi_{\psi}(k,t_{\rm e})}{M_{\rm P}}\biggr) m t,
\end{equation}
which is the same as Eq. (\ref{zetfina1}) since at $t_{\rm e}$, according to 
Eq. (\ref{incon1}), $\chi_{\psi}(k, t_{\rm e}) \equiv 
v_{\psi}(k, t_{\rm e})$.

\end{appendix}

\end{document}